\documentclass[twocolumn,showpacs,preprintnumbers,amsmath,amssymb,prl,superscriptaddress]{revtex4-1}
\usepackage{amsfonts}
\usepackage[english]{babel}
\usepackage[T1]{fontenc}
\usepackage{times}
\usepackage{mathrsfs}
\usepackage{graphicx}
\usepackage{dcolumn}
\usepackage{bm}
\usepackage[colorlinks,bookmarks=false,citecolor=blue,linkcolor=red,urlcolor=blue]{hyperref}

\newcommand\ket [1] {|#1 \rangle }
\newcommand\bra [1] {\langle #1 |}

\newcommand{\av}[1]{\langle #1\rangle}

\begin{document}

\title{Hierarchy of fractional Chern insulators and competing compressible states}

\author{A.M. L\"auchli}
\affiliation{Institut f\"ur Theoretische Physik, Universit\"at Innsbruck, A-6020 Innsbruck, Austria}
\affiliation{Max-Planck-Institut f\"ur Physik komplexer Systeme, N\"othnitzer Stra\ss e 38, D-01187 Dresden, Germany}
\author{Z. Liu}
\affiliation{Institute of Physics, Chinese Academy of Sciences, Beijing, 100190, China}
\author{E.J. Bergholtz}
\affiliation{Dahlem Center for Complex Quantum Systems and Institut f\"ur Theoretische Physik, Freie Universit\"at Berlin, Arnimallee 14, 14195 Berlin, Germany}
\affiliation{Max-Planck-Institut f\"ur Physik komplexer Systeme, N\"othnitzer Stra\ss e 38, D-01187 Dresden, Germany}
\author{R. Moessner}
\affiliation{Max-Planck-Institut f\"ur Physik komplexer Systeme, N\"othnitzer Stra\ss e 38, D-01187 Dresden, Germany}
\
\date{\today}

\begin{abstract}
We study the phase diagram of interacting electrons in a
dispersionless Chern band as a function of their filling. We find
hierarchy multiplets of incompressible states at fillings $\nu=1/3,
2/5, 3/7, 4/9, 5/9, 4/7, 3/5$ as well as $\nu=1/5,2/7$. These are accounted for by an
analogy to Haldane pseudopotentials extracted from an analysis of the
two-particle problem. Important distinctions to standard fractional
quantum Hall physics are striking: absent particle-hole symmetry in a
single band, an interaction-induced {\em single}-hole dispersion
appears, which perturbs and eventually destabilizes incompressible
states as $\nu$ increases. For this reason the nature of the state at 
$\nu=2/3$ is hard to pin down, while $\nu=5/7,4/5$ do not seem to 
be incompressible in our system.
\end{abstract}

\maketitle

{\em Introduction.---} Following recent proposals of the existence of novel lattice generalizations of
fractional quantum Hall (FQH) states, termed fractional Chern insulators (FCI), in 
(approximately) flat bands exhibiting non-zero Chern numbers~\cite{chernins1,chernins2,chernins3}, there
has been intense research activity in understanding this phenomenon~\cite{cherninsnum1,cherninsnum2,bosons,qi,bands,nonab1,nonab2,nonab3,cherncf,nthroot,others,WannierGaugeFixing,CompositeFCI,ThinTorusFCI,AdiabaticContinuity1,AdiabaticContinuity2,DipolarTFB,Grushin,ChernTwo,ChernN,disloc,interface,wannierpp}. 

From the original observation of a FCI state at Chern band 
filling $\nu=1/3$ \cite{chernins3,cherninsnum1,cherninsnum2} (at which the original FQH state was also first
observed), a number of questions immediately arise. Firstly, under
what conditions can FCIs be observed? Secondly, what are the states
which compete with the FCI states? Thirdly, what are the differences
between FCI physics in Chern bands compared to the familiar setting of
Landau levels in the continuum appropriate for describing the FQH
state in conventional semiconductor heterojunctions, e.g.\ arising due
to the non-uniform Berry curvature in reciprocal space?

This publication aims to contribute to all of these questions. We start
by demonstrating that a nearest neighbor interaction leads to FCI state also at $\nu=2/5$ and $\nu=1/5$, although the latter fraction has an FCI phase that is substantially less robust than the $\nu=1/3$ FCI. We account for this with a heuristic derived from considering the two-particle problem in the lattice model. Further, we find evidence
of several additional FCI states akin to the hierarchy FQH states familiar from conventional QH physics and its composite Fermion
 \cite{jain} hierarchy \cite{haldane83,halperin84} picture.

Turning to the qualitative distinctions from conventional QH physics,
we find that in the absence of particle-hole symmetry in a single band, an effective 
crystal-momentum--dependent potential appears, which shows up in the properties of the single-{\em hole} dispersion,
and leads to a modulation of the occupation numbers $n(\mathbf k)$ in the
many-body ground state which presage the breakdown of the QH
effect. This is reflected by a strong distinction between the
many-body states at $\nu=1/3$ and $\nu=2/3$ even in the spin-polarised
setting that we study. 

Our studies use extensive large-scale exact diagonalisations of the
many-body Hamiltonian of the lattice systems coupled with analytical
considerations. On a more technical level, we address questions of the
interplay of topological order and the concomitant finite-size lattice
quasi--ground-state degeneracies, as well as the finite-size scaling
of the gaps.

{\it Setup.---}
\begin{figure}[t]
\includegraphics[width=\linewidth]{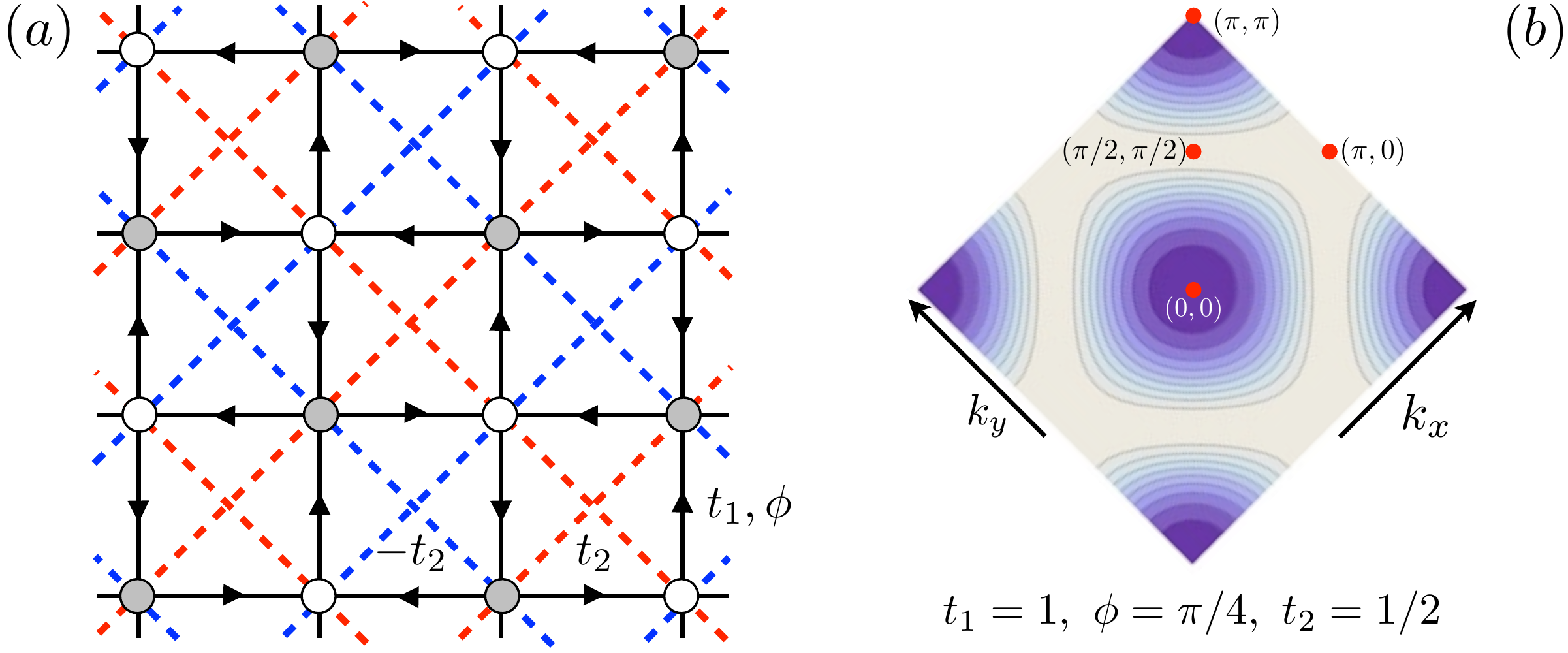}
\caption{(Color online)  (a) Illustration of the checkerboard lattice model with 
the relevant hopping amplitudes~\cite{chernins2,chernins3}. The interaction term
is a nearest neighbor density-density interaction (interaction between pairs of 
white and grey sites). (b) Berry curvature in the first Brillouin zone for
the indicated set of parameters. Dark (light) intensity denotes small (large) Berry curvature. 
} 
\label{fig:lattice_berry}
\end{figure} 
\begin{figure}[t]
\includegraphics[width=\linewidth]{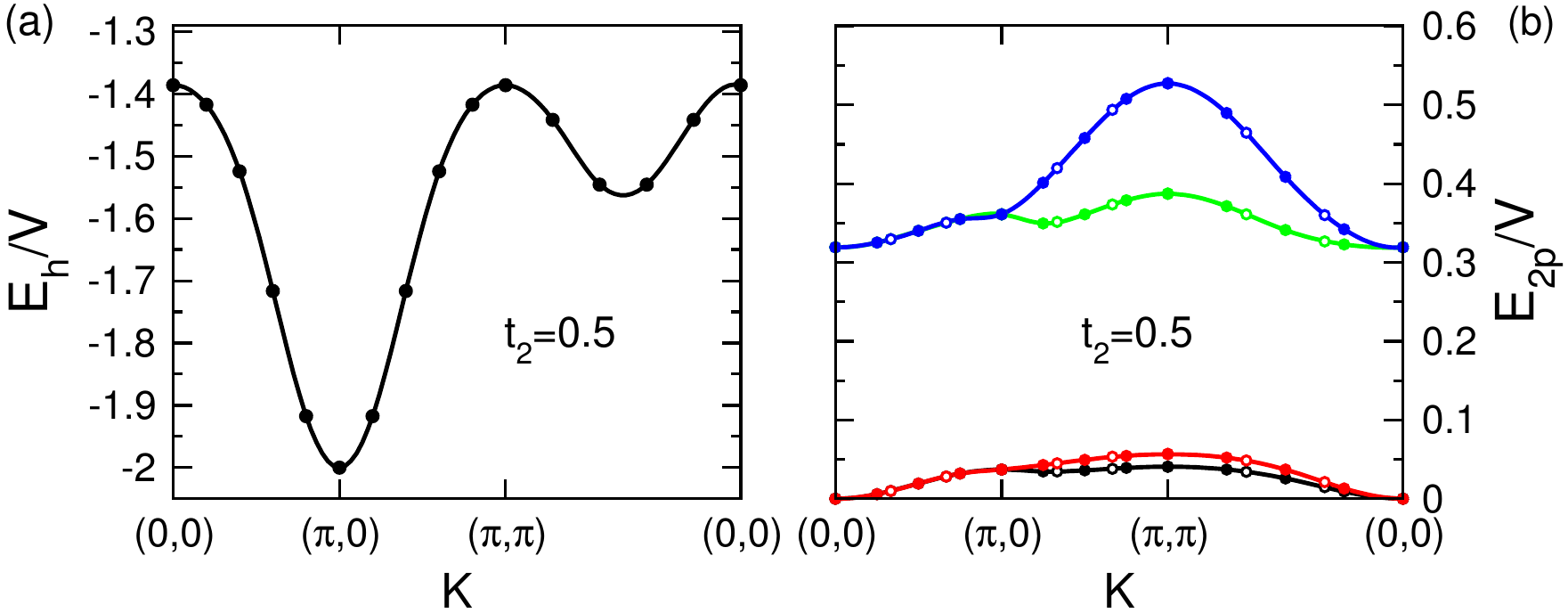}
\caption{(Color online) (a) Energy of a single hole in the fully filled lower band $E_h(\mathbf{K}):=E(N_p=N_s-1,\mathbf{K})-E(N_p=N_s,{\mathbf 0})$.
(b) Non-zero eigenvalues $E(N_p=2,\mathbf{K})$ of the two particle problem on a representative path through the Brillouin 
zone for $t_2=0.5$. In general there are four non-zero energies except at $\mathbf K = (0, 0)$ where there are only two finite energies. 
 Circles are data for (b) $6\times6$ (empty circles) 
and $8\times8$ (filled circles) unit cells [(a) $10\times10$ unit cells], and the lines are guides to the eye.} 
\label{fig:2P_1H}
\end{figure} 
\begin{figure*}[t]
\includegraphics[width=0.99\linewidth]{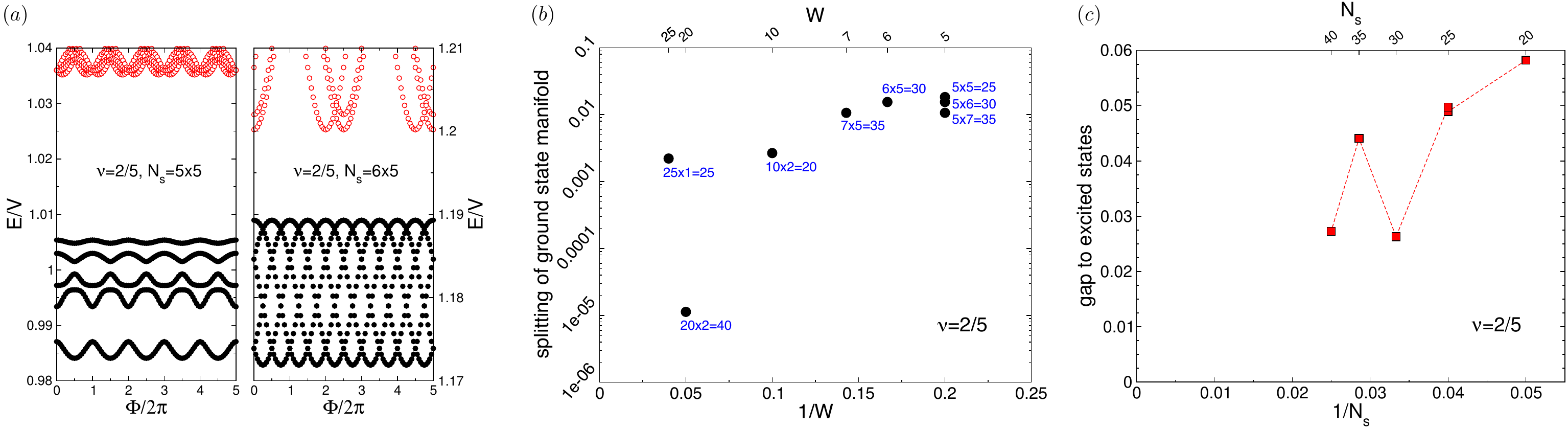}
\includegraphics[width=0.99\linewidth]{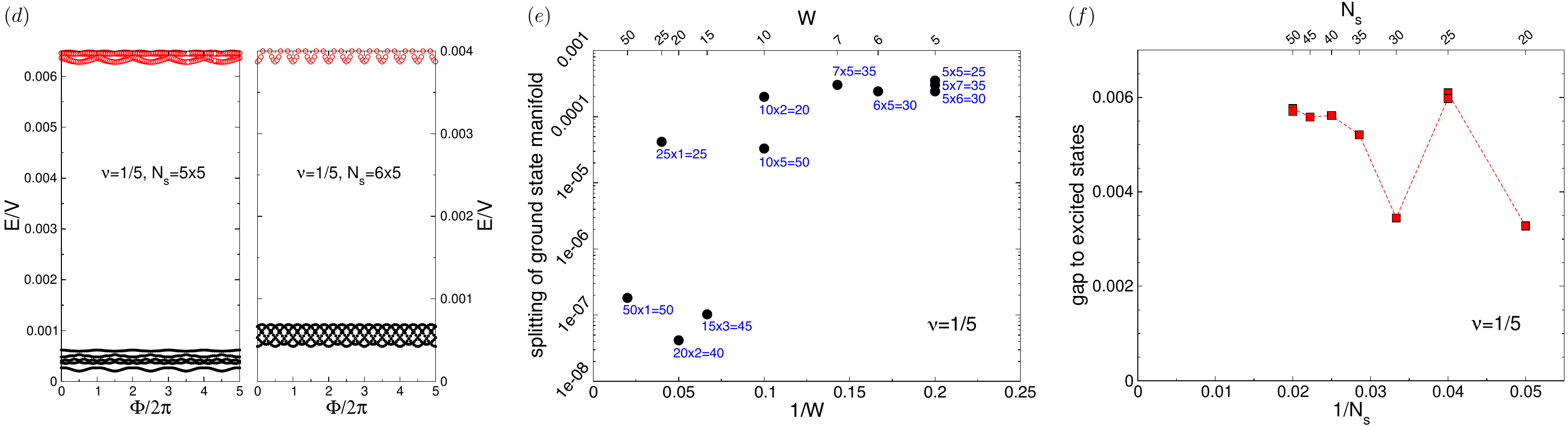}
\caption{(Color online) (a)-(c) Numerical evidence for a Fractional Chern insulator state at $\nu=2/5$. For details see main text. (d)-(f) Numerical evidence for a Fractional Chern insulator state at $\nu=1/5$.
Here we focus on $t_2=1/2$ which is close to the value where the variance of the Berry curvature is minimal. 
} \label{fig:numerics_p_1+2_q_5}
\end{figure*} 
In order to be specific we focus on the two-band checkerboard lattice model introduced in Refs.~\cite{chernins2,chernins3} and 
sketched in Fig.~\ref{fig:lattice_berry}(a). Here $t_1e^{\pm i\phi}$ is a nearest-neighbor hopping with an orientation dependent 
complex phase, and $t_2$ denotes the next-nearest-neighbor hopping amplitude.
After Fourier transform, the single-electron (kinetic) Hamiltonian reads
\begin{eqnarray}
\label{checkerboard}
\mathcal{H}=\sum_{\mathbf k\in \textrm{BZ}}(c^{\dagger}_{\mathbf kA},c^{\dagger}_{\mathbf kB})
h(\mathbf k)(c_{\mathbf kA},c_{\mathbf kB})^T,\label{tb}
\end{eqnarray}
where $h(\mathbf k)$ is given as~\cite{sp_hamilton}.
In the following we consider the case of $t_1=1, \phi=\pi/4$ and varying $t_2$. For $t_2\neq 0,\infty$ one obtains two separated bands
with Chern number $\pm1$.
The Berry curvature in each band can be expressed in terms of the single-particle states in the band, $\ket{n_{\mathbf k}^\pm}$, via the Berry connection 
$A_j^\pm({\mathbf k})=-i\bra{n_{\mathbf k}^\pm}\partial_{k_j}\ket{n_{\mathbf k}^\pm}$ as $F_{ij}^\pm=\partial_{k_i}A_j^\pm({\mathbf k})-\partial_{k_j}A_i^\pm({\mathbf k})$. 
The two bands have topologically quantized Chern numbers, $C=\frac 1 {2\pi}\int_{BZ} F_{12}^\pm({\mathbf k})d^2k=\pm1$, while the Berry curvature varies in the Brillouin zone and depends on the microscopic parameters ($t_2$ in the present case). In Fig.~\ref{fig:lattice_berry}(b) we display the Berry curvature for the specific value of $t_2=0.5$ [note that the Berry curvature
vanishes at $(0,0)$ and $(\pi,\pi)$].

\paragraph{Interactions.---}

We consider nearest neighbor repulsion, $H_\textrm{int}=V\sum_{\av{i,j}}n_in_j$ [cf. caption of Fig.~\ref{fig:lattice_berry}(a)], throughout this publication.
If the energy gap between the two bands is large compared to the interaction strength it is possible to project the interactions into the partially filled band, 
leading to a projected Hamiltonian of the form 
\begin{eqnarray}H=\sum_{{\mathbf k}_1
 \mathbf k_2 \mathbf k_3 \mathbf k_4}\!\!\!
 V_{\mathbf k_1 \mathbf k_2 \mathbf k_3 \mathbf k_4}c^{\dagger}_{\mathbf k_1}c^{\dagger}_{\mathbf k_2}c_{\mathbf k_3}c_{\mathbf k_4} ,\label{hamf}\end{eqnarray}
for two-body interactions. Here, $c_{\mathbf{Q}}$ annihilates an electron with wavefunction $\phi_{\mathbf{Q}} (\mathbf{r})$ of the single particle state of the partially filled band
with crystal momentum $\mathbf{Q}$. In the following we take the flat band limit, i.e. drop the residual energy dispersion in the band, as we wish to highlight the new phenomena
arising due to the non-constant Berry curvature, as opposed to remnant energy dispersion.

{\it Particle-hole asymmetry and two-particle analysis.---}
It is enlightening to consider the two simplest interacting cases, namely the single-hole and the two-particle problem. 
Performing a particle-hole transformation, $c_\mathbf k \rightarrow c_\mathbf k^\dagger$ within one of the bands, the projected 
Hamiltonian (\ref{hamf}) transforms to 
\begin{eqnarray}H\rightarrow \sum_{\mathbf k_1
 \mathbf k_2 \mathbf k_3 \mathbf k_4}\!\!\!
 V_{\mathbf k_1 \mathbf k_2 \mathbf k_3 \mathbf k_4}^*c^{\dagger}_{\mathbf k_1}c^{\dagger}_{\mathbf k_2}c_{\mathbf k_3}c_{\mathbf k_4} +\sum_{\mathbf k}E_h(\mathbf k)
c^{\dagger}_{\mathbf k}c_{\mathbf k},\label{ph}\end{eqnarray}
which includes an effective single-hole energy $E_h(\mathbf k)=4\sum_{\mathbf m}
 V_{\mathbf m \mathbf k \mathbf m \mathbf k}$ (adopting a convention under which $V_{\mathbf{klmn}}$ is antisymmetric under exchange of its indices). This term amounts to a trivial overall energy shift in a Landau level. By contrast, it introduces an effective dispersion {\em even for an entirely flat} Chern band! We display $E_{h}(\mathbf{k})$ in Fig.~\ref{fig:2P_1H}(a) for $t_2=0.5$. As we will demonstrate below, this particle-hole asymmetry generally leads to a deformation of $n(\mathbf k)$ and in fact dominates the physics near $\nu=1$, i.e. in the dilute-hole limit. Note that for our checkerboard example, the energy
spectrum is particle-hole invariant under $\nu \leftrightarrow 2 - \nu$. 

Next, we consider the two particle problem: In Fig.~\ref{fig:2P_1H}(b) we display the non-zero eigenvalues of the two particle problem along a path in the Brillouin zone
for the nearest neighbor interaction for $t_2=0.5$. The spectrum depends on the total momentum, $\mathbf K=\mathbf k_1+\mathbf k_2$, underscoring the lack of translation invariance in reciprocal space. We observe two dominant eigenvalues of mean value $\approx 0.4V$ which remain non zero throughout the Brillouin zone, while there is also second set of two eigenvalues about $10$ times smaller which vanish as $\mathbf K \rightarrow (0, 0)$. All other eigenvalues are strictly zero. Quite generally, for each $\mathbf{K}$, the number of non-zero energy levels is bounded above by the number of finite energy levels of the interaction alone before band projection. This follows from the fact that each (unprojected) interaction term imposes one (linear) constraint which needs to be satisfied for the two particle wavefunction $| \psi_{\mathbf K} \rangle = \sum_{\mathbf q} \alpha^\mathbf{K}_{\mathbf q} \left[ \phi_{{\mathbf K} + {\mathbf q}} (r_1) \phi_{-\mathbf q} (r_2) - \phi_{{\mathbf K} + {\mathbf q}} (r_2) \phi_{-\mathbf q} (r_1) \right]$ to have zero energy: $\langle \psi_{\mathbf K} | V (r_1, r_2) | \psi_{\mathbf K} \rangle = 0$. If these constraints are linearly dependent, there can be additional zero-energy states, as is the case at $\mathbf K =(0, 0)$ in the present example. Here, the upper bound of $4$ finite eigenvalues is saturated everywhere else in the Brillouin zone. Including also a next-nearest neighbor repulsion leads to $8$ finite eigenvalues, and e.g.~on a three-sublattice kagome lattice model \cite{chernins1} one finds $6$ (12) levels including the (up to next-)nearest neighbor terms. 

These observations motivate a suggestive analogy between the finite energy levels of the projected two-particle problem and Haldane's pseudopotenials well known in the continuum Landau levels \cite{haldane83}. We thus tentatively label the two non-zero pairs of eigenvalues as ``pseudopotentials'' $\mathcal V_ 1$ and $\mathcal V_3$.

{\it Hierarchy states and the phase diagram.---} To investigate the full interacting many-body problem we have performed extensive exact diagonalization studies on a large number of finite samples with rectangular shapes 
(as in previous numerical studies of FCI) as well as tilted samples, where the spanning vectors of the samples need not be aligned with the lattice axes. The latter choice of samples allows to study a considerably larger number of clusters with aspect ratio close to one. The list of considered clusters can
be found in the supplementary material~\cite{suppmat}.

The pseudopotential analogy introduced above suggests a hierarchy of incompressible states at $\nu=p/q\geq 1/3$, $q$ odd, due to the  relatively large energy scale $\mathcal V_1$. In particular, we have looked for the $m = 1$ composite fermion branch at $\nu=p/(2mp+1)$~\cite{jain} with rather clear evidence of FCI states at $\nu=2/5,3/7,4/9$ as well as their $1-\nu$ relatives at $\nu=5/9,4/7,3/5$, 
beyond the $\nu=1/3$ state already discussed in the literature~\cite{chernins3,cherninsnum1,cherninsnum2}. Data of the low-energy spectrum of at least two samples for each fraction are displayed in the supplementary material~\cite{suppmat}. Here we provide strong evidence for a stable incompressible $\nu=2/5$ state as shown in Fig.~\ref{fig:numerics_p_1+2_q_5}(a)-(c). In panel Fig.~\ref{fig:numerics_p_1+2_q_5}(a) we present the spectral flow of the 5 ground state manifold levels upon flux insertion for two different samples. 
The ground state manifold (filled symbols)  does not mix with  excited states (empty symbols) and returns to the initial spectrum upon insertion of five flux quanta. This provides strong evidence for 
a state with quantized Hall conductance. In Fig.~\ref{fig:numerics_p_1+2_q_5}(b) we display the finite size scaling of the energy spread of the five states in the ground state manifold for  different samples. We plot the energy splitting as a function of the inverse topological diameter $1/W$~(as defined in the supplementary material~\cite{suppmat}), which can become quite small (as small as $1/25$) for our tilted samples. Note that the topological extent is not in general equivalent to the geometrical extent. In particular the samples with $W=10,20,25$ have a geometrical aspect ratio equal to one. We also stress that
the five ground states are systematically found in the sectors predicted by the counting rule developed in Refs.~\cite{cherninsnum2,nonab1}, and which we generalized for tilted samples~\cite{suppmat}. 
In the case of tilted samples the ground state sectors show a great variability depending on the cluster geometry, which is an argument against the formation of a charge density wave state, which should exhibit 
a unique set of degenerate momenta dictated by the spatial symmetry breaking of the charge density wave. Finally, in  Fig.~\ref{fig:numerics_p_1+2_q_5}(c) we show the finite size scaling of the gap from the ground state to the first excited state as a function of inverse systems size $1/N_s$. Our data shows non-negligible 
finite size effects, but convincingly points towards a finite excitation gap of the order of $V/30$ in the thermodynamic limit, in contrast to recent claims denying a stable $\nu=2/5$ state for the checkerboard lattice~\cite{CompositeFCI,AdiabaticContinuity2}.

Next we turn to fractions below $\nu=1/3$. According to the pseudopotential analogy our nearest neighbor interactions generate a finite $\mathcal{V}_3$, 
albeit roughly a factor ten smaller than $\mathcal{V}_1$. We would
thus expect to find a stable $\nu=1/5$ state with associated energy scales significantly smaller than those of the $\nu=1/3$ state. Indeed we find strong evidence for a $\nu=1/5$ FCI state as shown in Fig.~\ref{fig:numerics_p_1+2_q_5}(d)-(f). In particular the finite size effects of the ground state manifold splitting shown in Fig.~\ref{fig:numerics_p_1+2_q_5}(e) behave qualitatively similar to the $\nu=2/5$ case in Fig.~\ref{fig:numerics_p_1+2_q_5}(b), and the energy gap to excited states extrapolates very nicely to a value of the order of $V/200$. We furthermore find some mild evidence for a $\nu=2/7$ state~\cite{suppmat}, which is stabilized by $\mathcal{V}_3$.

\begin{figure}[ht]
\centerline{\includegraphics[width=0.99\linewidth]{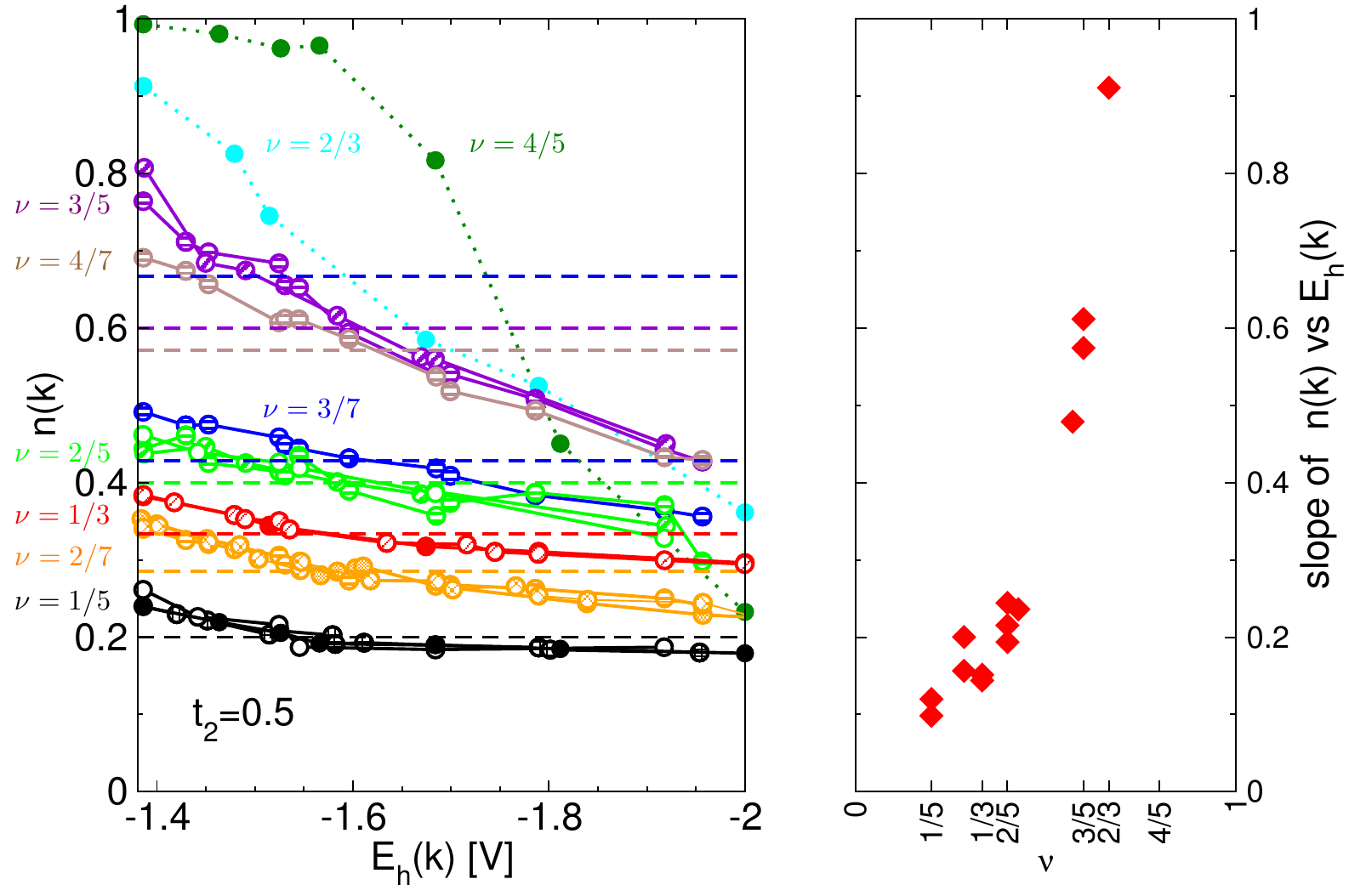}}
\caption{(Color online) left panel: Momentum space orbital occupation $n(\mathbf k)=\langle c^{\dagger}_{\mathbf k} c_{\mathbf k}\rangle$ 
plotted as a function of the single-hole energy $E_h(\mathbf k)$. As the filling $\nu$ is increased, $n(\mathbf{k})$ displays a more pronounced
(close to linear) correlation with $E_h(\mathbf k)$. Dashed lines denote fractions which are most likely not FCI states. 
Right panel: Approximate slope of the correlation between $n(\mathbf{k})$ and $E_h(\mathbf k)$ plotted as
a function of the filling $\nu$. The rapid, nonlinear increase of the slope for large fillings is a likely cause for the breakdown of FCI states at large filling.
}
\label{fig:Eh_vs_nk}
\end{figure} 

\paragraph{Effect of the single hole energy. ---}

When numerically exploring the phase diagram we discovered that several particle hole conjugate states of stable low-density FCI fractions do not seem to be realized, for example
we find it difficult to systematically observe the required ground state degeneracy for $\nu=2/3$, while for $\nu=5/7$ and $\nu=4/5$ the required degeneracy is absent.
An enlightening  way to understand this finding is to monitor the filling dependence of the momentum space
occupation number $n(\mathbf k)=\langle c^{\dagger}_{\mathbf k} c_{\mathbf k}\rangle$ when plotted as a function of the single hole energy $E_h(\mathbf k)$ introduced before, see Fig.~\ref{fig:Eh_vs_nk}(a),
 where the occupancy $n({\mathbf k})$ in the incompressible many-body state clearly tracks $E_h({\mathbf k})$. The role of the interaction-induced effective single particle dispersion
is thus to significantly distort the momentum space occupancy as the filling increases, and ultimately leads to Fermi-surface like compressible states which are governed by the effective single particle dispersion. This behavior is illustrated
in Fig.~\ref{fig:Eh_vs_nk}(b) with the drastic increase of the correlation between $n(\mathbf k)$ and $E_h({\mathbf k})$ as $\nu$ increases, and the Fermi-surface like $n(\mathbf k)$ in the
$\nu=4/5$ state depicted in Fig.~\ref{fig:Eh_vs_nk}(a).

It is instructive to compare our results from what is known from conventional FQH physics.
For intermediate filling fractions in the lowest Landau level, both experiment \cite{pan} and theory \cite{hierarchy} suggest that the stability  of the FQH states towards disorder, temperature etc., is essentially determined by the denominator, $q$, of the filling fraction (for odd $q$) implying a self-similar structure in the clean zero-temperature limit. Corrections to the self-similar structure in "real" systems are minor and include effects due to an instability towards Wigner crystallization at low filling fractions and particle-hole symmetry breaking effects (LL mixing etc). The main features are qualitatively similar to what we find in the lattice system. However, in the Chern insulator case the corrections are more substantial. In particular, there is nothing like a "clean" limit as the Berry curvature effects necessarily destroys weak states, and the compressible competing states discussed above prevail. Moreover, due to the explicit particle-hole asymmetry  within the band we find that FCI states are absent for $\nu\gtrsim 2/3$. 

\paragraph{Discussion.---} 
The question to what extent fractional Chern insulators are `nothing but' FQH states on the lattice is fundamental to this rapidly developing field: it combines the questions of where to find a FCI with how to diagnose it and how to theoretically describe it, before being able to address differences to the `weak-field' continuum Landau level physics.

The FCI states we thus found, as well as their respective stabilities, are naturally organised with the aid of the hierarchy concept developed for the FQH. To obtain this information, we have developed new diagnostics, including an adaptation of Haldane's pseudopotentials, strongly suggesting a common field-theoretic description for the two phenomena. At the same time, a particularly noteworthy {\em negative} result is our observation, at odds with some previous treatments, that FCIs do not generically come with a constant occupation number of the flat band orbitals in reciprocal space.

Indeed, this phenomenon arises even for perfectly flat (dispersionless) bands, as a result of a non-uniform Berry curvature, which appears as a potent driving force for the destruction of FCI states by acting as a $k$-dependent chemical potential of a strength tuned by the electron-density of the flat band, removing particle-hole symmetry in a lattice-specific manner. Our study provides a  a first glimpse of the competing states. 

Clearly, much further work on each of these points is warranted, especially in the light of promising experimental prospects to observe such states in optical lattices with artificial magnetic fields~\cite{HofstadterOpticalLattice,HofstadterBerry,FluxLatticeBerry,DipolarTFB}.

\paragraph{Note added.---} While completing the present work we became aware of two preprints which also provide evidence for 
FCI states beyond the Laughlin states~\cite{CompositeFCI,AdiabaticContinuity2}.

\acknowledgments

We acknowledge useful discussions with Masud Haque. EJB is supported by the Alexander von Humboldt foundation.
Simulations have been performed on machines of the platform "Scientific computing" at the University of
Innsbruck - supported by the BMWF - and on the PKS-AIMS and vip cluster at the MPG RZ Garching.

\newpage
\section*{Supplementary material}
\begin{figure}
\includegraphics[width=\linewidth]{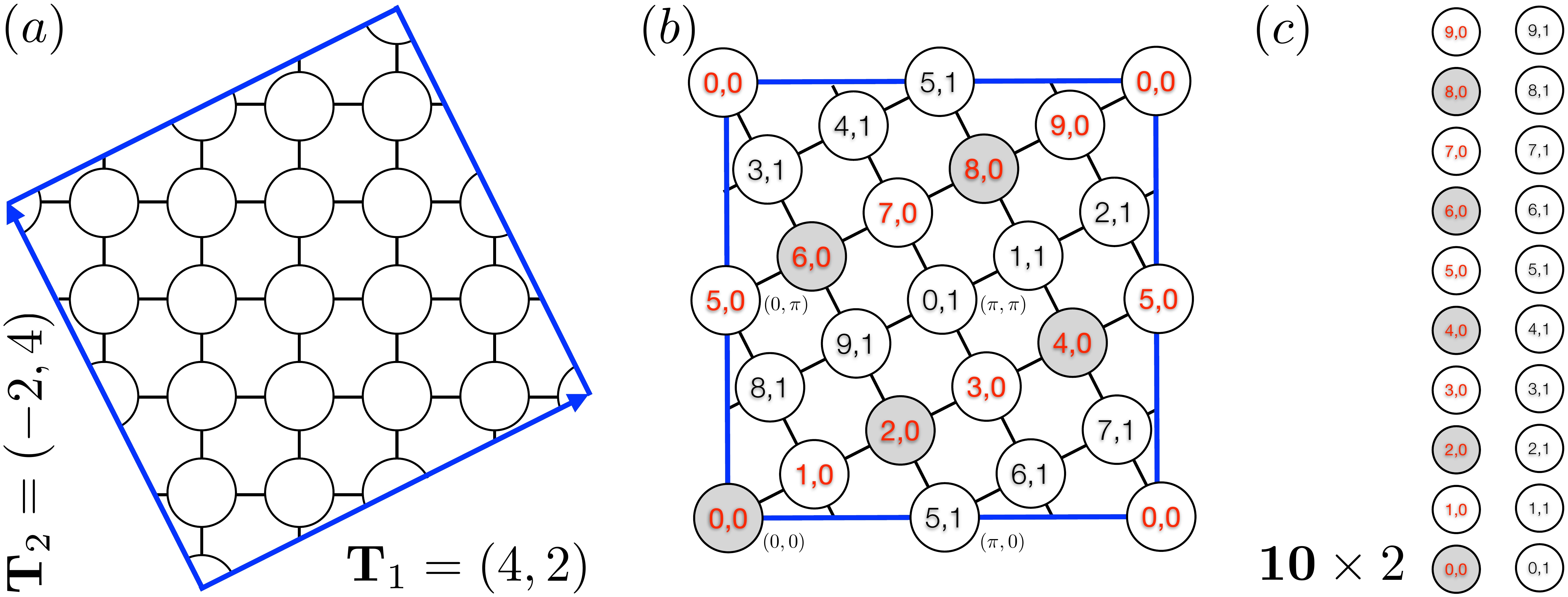}
\caption{(Color online) Geometry of a $N_s=20$ site 
cluster with the spanning vectors $\mathbf{T}_1$ and $ \mathbf{T}_2$,
its momentum space point and an illustration of the topological extent.
The shaded grey momentum points correspond to the expected 
ground state momenta for $\nu=1/5$ state derived by the counting
rule for a $10\times 2$ lattice.
} 
\label{fig:tiltedcluster}
\end{figure} 

\begin{table}[t]
\begin{tabular}{|c|r|r|}
\hline
$N_s$&Spanning vectors& Topological extent  \\
&$\mathbf{T}_1 / \mathbf{T}_2$&$W \times U$\\
\hline
18 &$(3,3)/(-3,3)$ &  $\mathbf{6}\times3$\\
20 &$(4,2)/(-2,4)$ &  $\mathbf{10}\times2$\\
25 &$(5,0)/(0,5)$ &  $\mathbf{5}\times \mathbf{5}$\\
25 &$(4,3)/(-3,4)$ &  $\mathbf{25}\times1$\\
28 &$(4,0)/(0,7)$ &  $\mathbf{4}\times \mathbf{7}$\\
30 &$(5,0)/(0,6)$ &  $\mathbf{5}\times \mathbf{6}$\\
35 &$(5,0)/(0,7)$ &  $\mathbf{5}\times \mathbf{7}$\\
36 &$(6,0)/(0,6)$ &  $\mathbf{6}\times \mathbf{6}$\\
40 &$(6,2)/(-2,6)$ &  $\mathbf{20}\times 2$\\
45 &$(6,3)/(-3,6)$ &  $\mathbf{15}\times 3$\\
50 &$(5,5)/(-5,5)$ &  $\mathbf{10}\times 5$\\
50 &$(7,1)/(-1,7)$ &  $\mathbf{50}\times 1$\\
\hline
\end{tabular}
\caption{List of samples studied in the present work. Apart from
the rectangular samples with $28,30$ and $35$ sites, all  clusters
have an geometric aspect ratio of 1 and at least a $C_4$ point group symmetry
in addition to the translation group.
}
\label{tab:samples}
\end{table}

\subsection*{Tilted Square samples}

We use rectangular and tilted samples in our exact diagonalization study.
The samples are defined by their spanning vectors $\mathbf{T}_1$ and $ \mathbf{T}_2$,
which define the toroidal periodic boundary conditions. A list of clusters studied in this
work is shown in Tab.~\ref{tab:samples}. The tilted clusters used here all have perpendicular
spanning vectors $\mathbf{T}_1$ and $ \mathbf{T}_2$ of equal length, but these requirements 
could also be relaxed in future studies for an even larger sample variability. An interesting aspect
of the tilted samples is their so called {\em topological} extent. While they are symmetric samples
of aspect ratio one in real space [cf. Fig.~\ref{fig:tiltedcluster}(a)], they actually have a different 
aspect ratio in momentum space. We define the topological extent $W$ as the number of steps 
one has to go in momentum space on a straight path along nearest neighbor orbitals in 
momentum space until one returns to the starting point. The complementary extent $U$ is then
simply defined as $U=N_s/W$. The topological extent of all clusters is listed in Tab.~~\ref{tab:samples}.
The extents $W$ and $U$ are important for two reasons: i) it seems that the parameter $W$ controls 
to a large extent the splitting of the ground state manifold
as demonstrated for $\nu=1/5$ and $\nu=2/5$ in Fig.~\ref{fig:numerics_p_1+2_q_5}(b),(e) of the main
paper. This then allows to probe for the existence of a degenerate ground state manifold on much smaller
clusters compared to the usually employed rectangular clusters, where the topological extent is identical to
the real space extent, and thus values for $W$ of the order of ten or more will be difficult to reach with exact diagonalization.
The second reason ii) is that the counting rule for ground state quantum number prediction~\cite{cherninsnum2,nonab1}
can then be applied to the topological $W\times U$ lattice, and then be embedded back into the original momentum space
to obtain the expected ground state momenta at a certain filling. This is illustrated by the shaded circles in Fig.~\ref{fig:tiltedcluster}
(b) and (c) for a specific filling fraction.

\subsection*{Numerical support for hierarchy states}

In the following we present low energy spectra for various fraction where
there is some evidence for FCI states. These are the fractions $\nu=1/5,\mathbf{2/7},1/3,\mathbf{2/5, 3/7, 4/9, 5/9, 4/7, 3/5}$.
We find the situation at $\nu=\mathbf{2/3}$ to be unclear, while $\nu=\mathbf{5/7}$ and $\nu=\mathbf{4/5}$ do not show 
any evidence for FCI states. For the bold fractions we now show the low energy spectrum for two or more samples. 
For the FCI fractions the ground state exhibits the correct approximate degeneracy
at the momenta predicted by the counting rule~\cite{cherninsnum2,nonab1} in some range of $t_2$ values.

Apart from the fraction $\nu=1/5$ and $\nu=2/5$ discussed in the main paper 
(as well as $\nu=1/3$ discussed in the literature~\cite{cherninsnum1,chernins3,cherninsnum2}) it is quite difficult to 
perform a systematic finite size scaling study, because of the sometimes 
pronounced sample dependent results, as well as due to limited sample availability
for certain fractions. In all of the following plots the expected ground state sectors 
are denoted by the filled symbols in different colors, whereas the other sectors are
denoted by hashed symbols. The range of $t_2$ values where
the correct approximate ground state degeneracy can be found before the first 
excited state is denoted by an orange box in each graph. As the data is shown
it is in principle only evidence for the emergent many-body translation symmetry, similar to the
FQH states on the torus, where {\em any} state at  filling $\nu=p/q$ is $q$-fold degenerate.
We however also checked the flux threading property of the ground state manifold -- similar
to Fig~\ref{fig:numerics_p_1+2_q_5}(a),(d) -- for several of the fractions, with a positive results.
One could also perform quasihole counting, but this requires care
because of aliasing problems, for example on the $N_s=35$ site sample
the states at $\nu=2/5$ and $\nu=3/7$ (and at $\nu=3/5$ and $\nu=4/7$) just differ by one electron.

\begin{figure*}
\includegraphics[width=\linewidth]{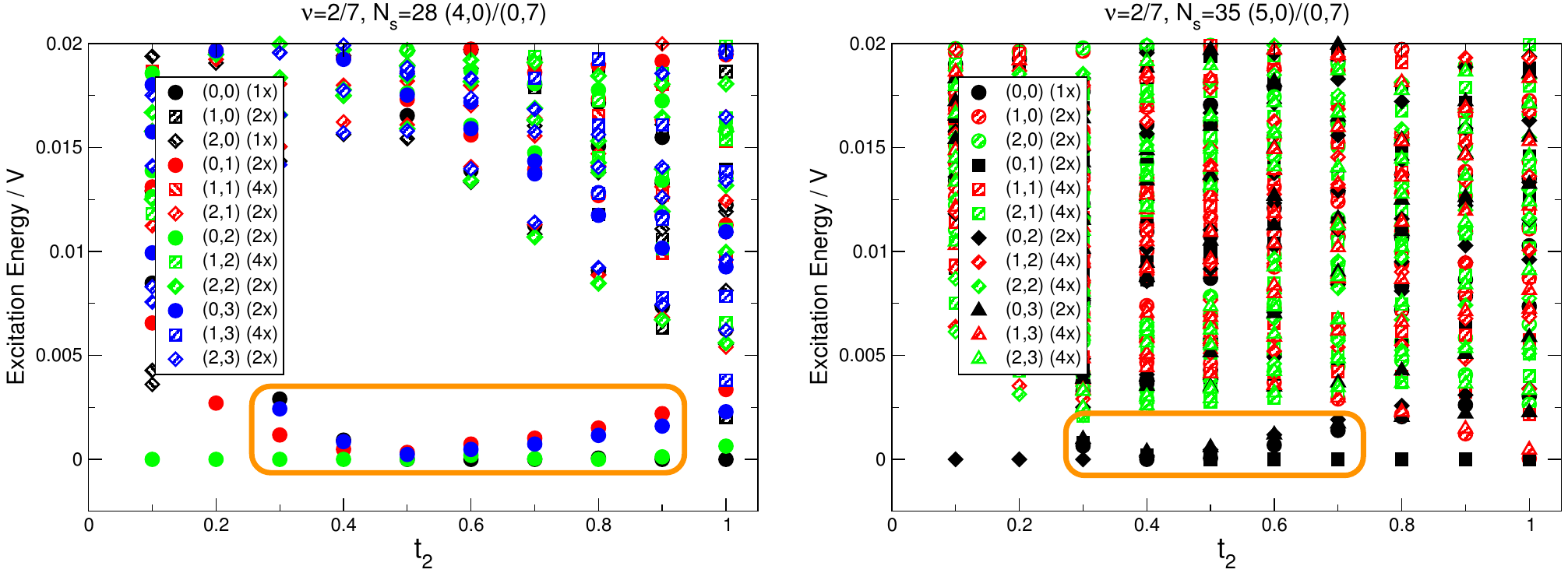}
\caption{(Color online) 
FCI at $\nu=2/7$
}
\label{fig:supp_p_2_q_7}
\end{figure*}

\begin{figure*}
\includegraphics[width=\linewidth]{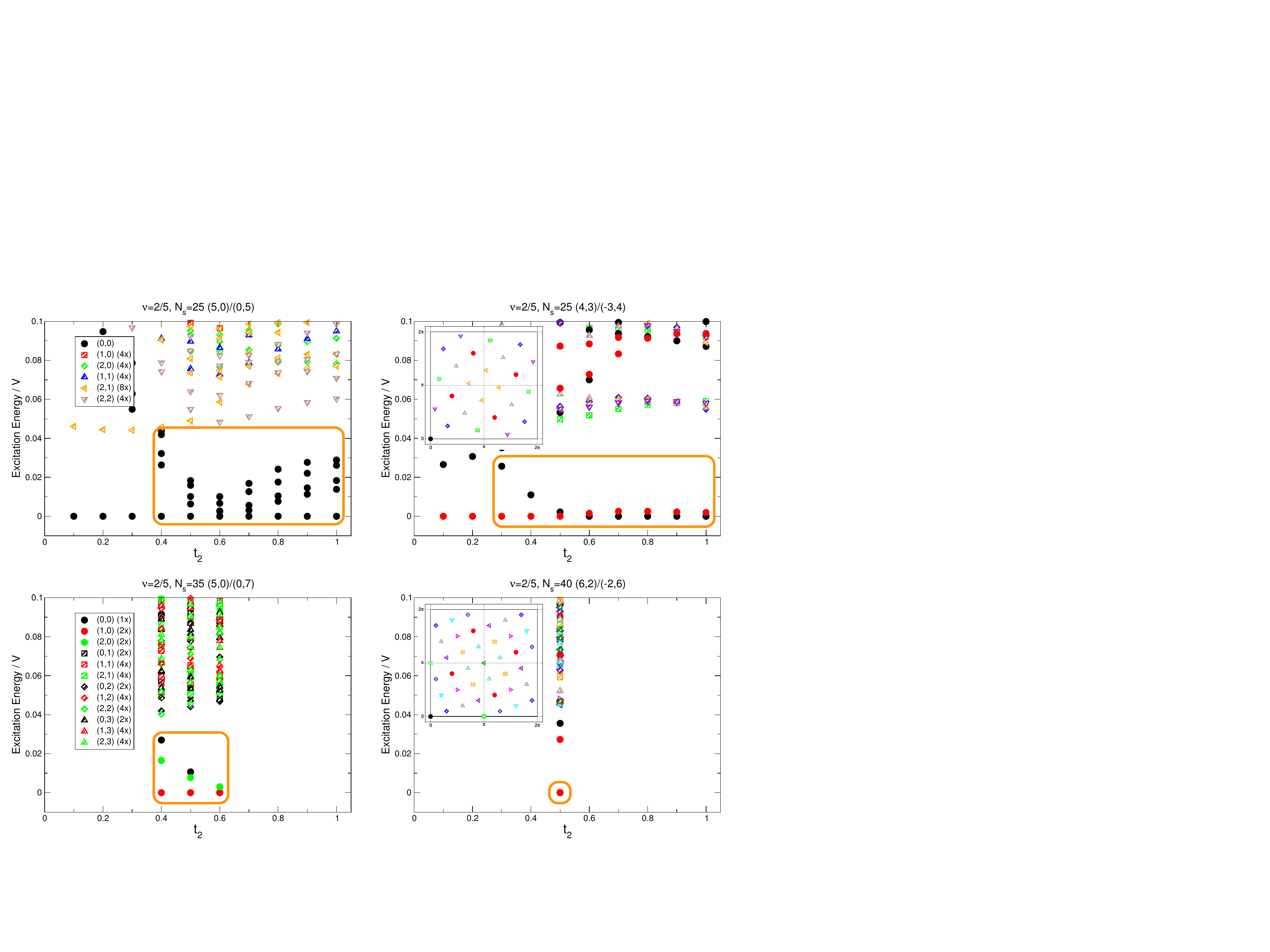}
\caption{(Color online) 
FCI at $\nu=2/5$
}
\label{fig:supp_p_2_q_5}
\end{figure*}
\begin{figure*}
\includegraphics[width=\linewidth]{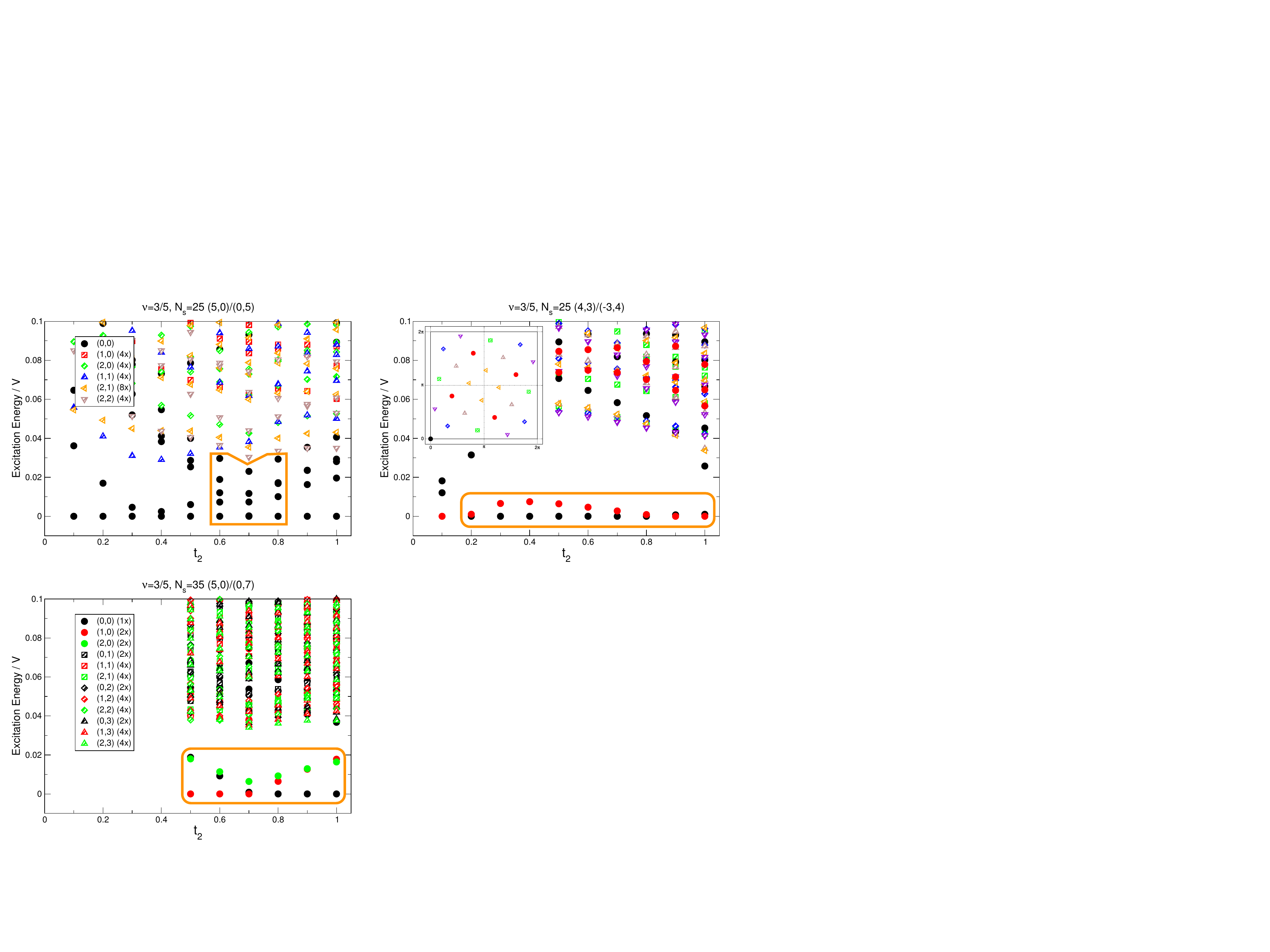}
\caption{(Color online) 
FCI at $\nu=3/5$ 
}
\label{fig:supp_p_3_q_5}
\end{figure*}

\begin{figure*}
\includegraphics[width=\linewidth]{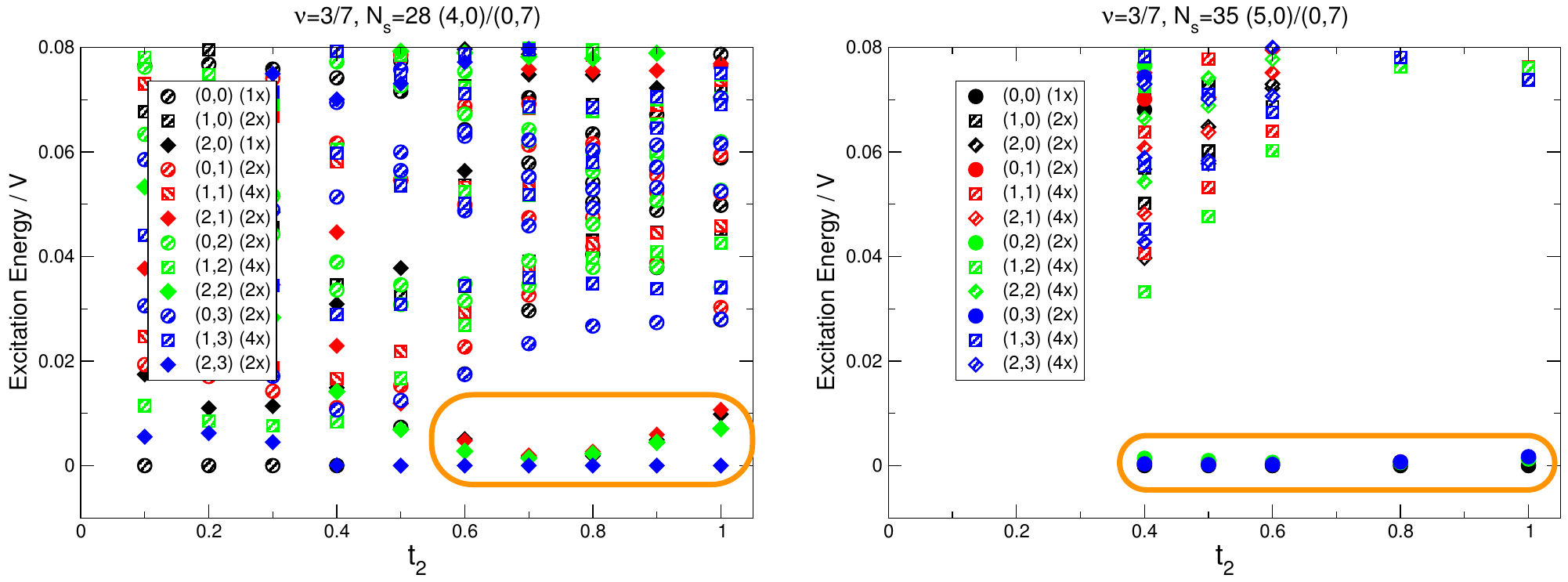}
\includegraphics[width=\linewidth]{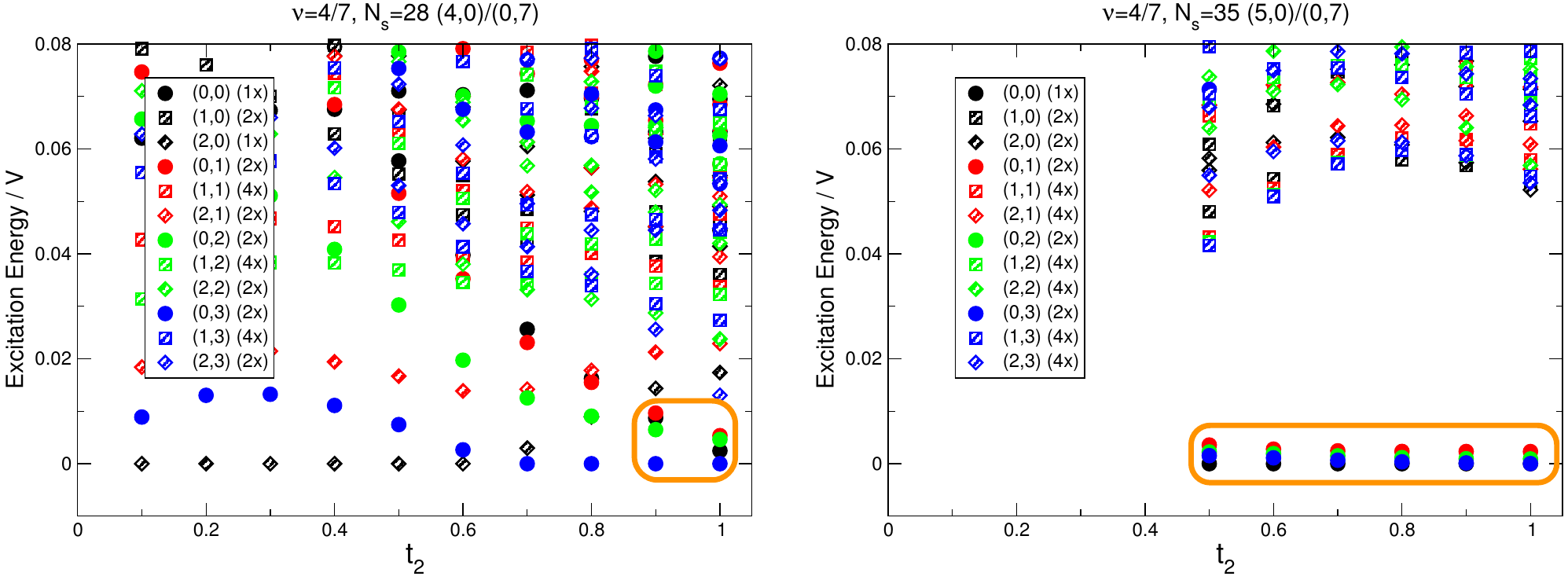}
\caption{(Color online) 
FCI at $\nu=3/7$ and $\nu=4/7$
}
\label{fig:supp_q_7}
\end{figure*}

\begin{figure*}
\includegraphics[width=\linewidth]{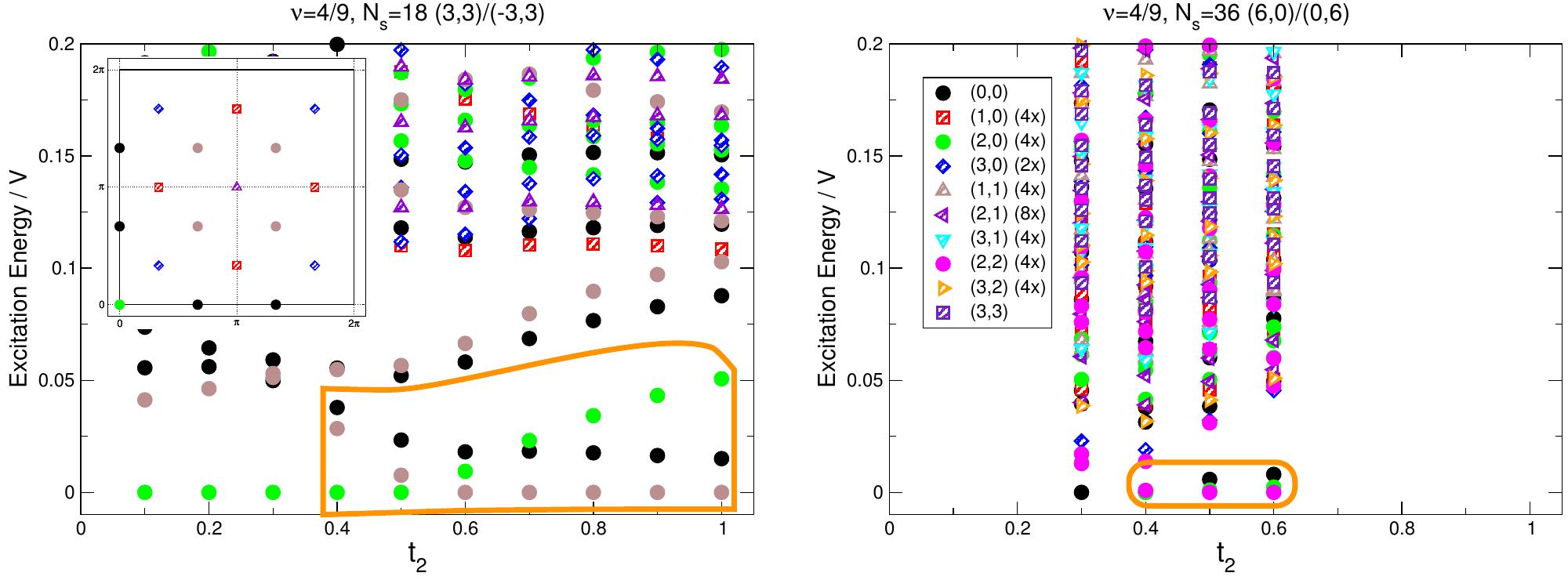}
\includegraphics[width=\linewidth]{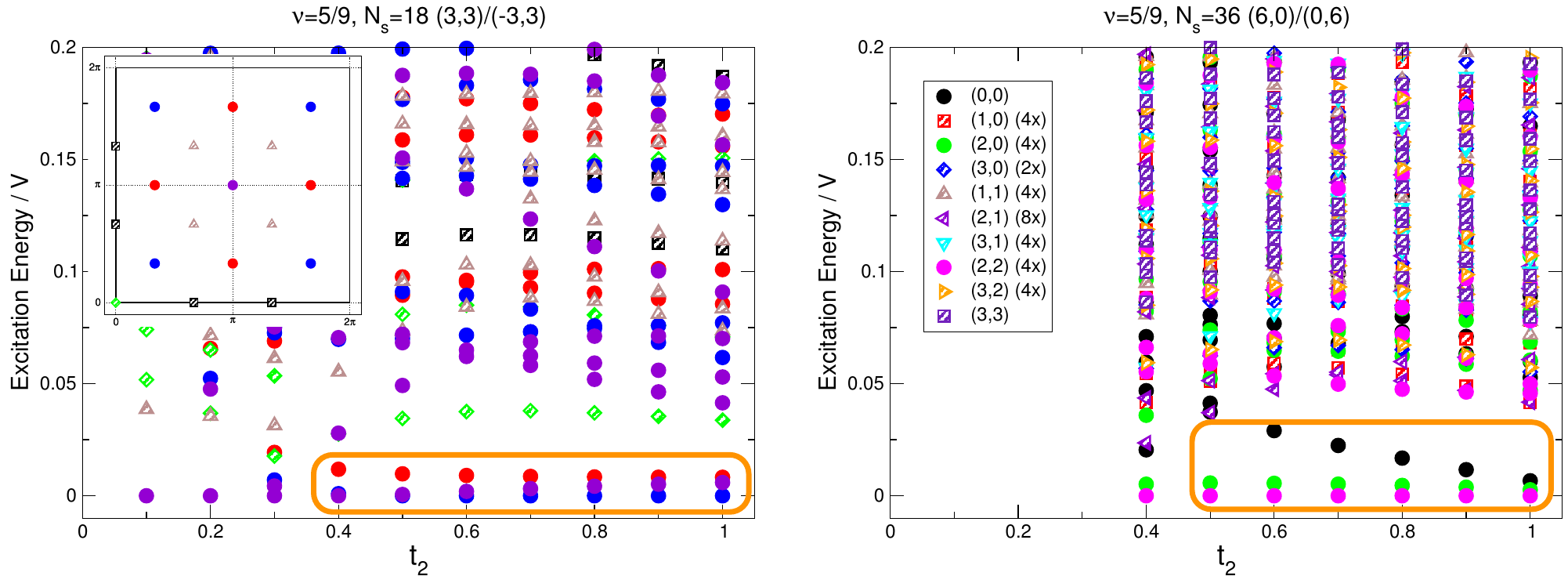}
\caption{(Color online) 
FCI at $\nu=4/9$ and $\nu=5/9$
}
\label{fig:supp_q_9}
\end{figure*}

\begin{figure*}
\includegraphics[width=\linewidth]{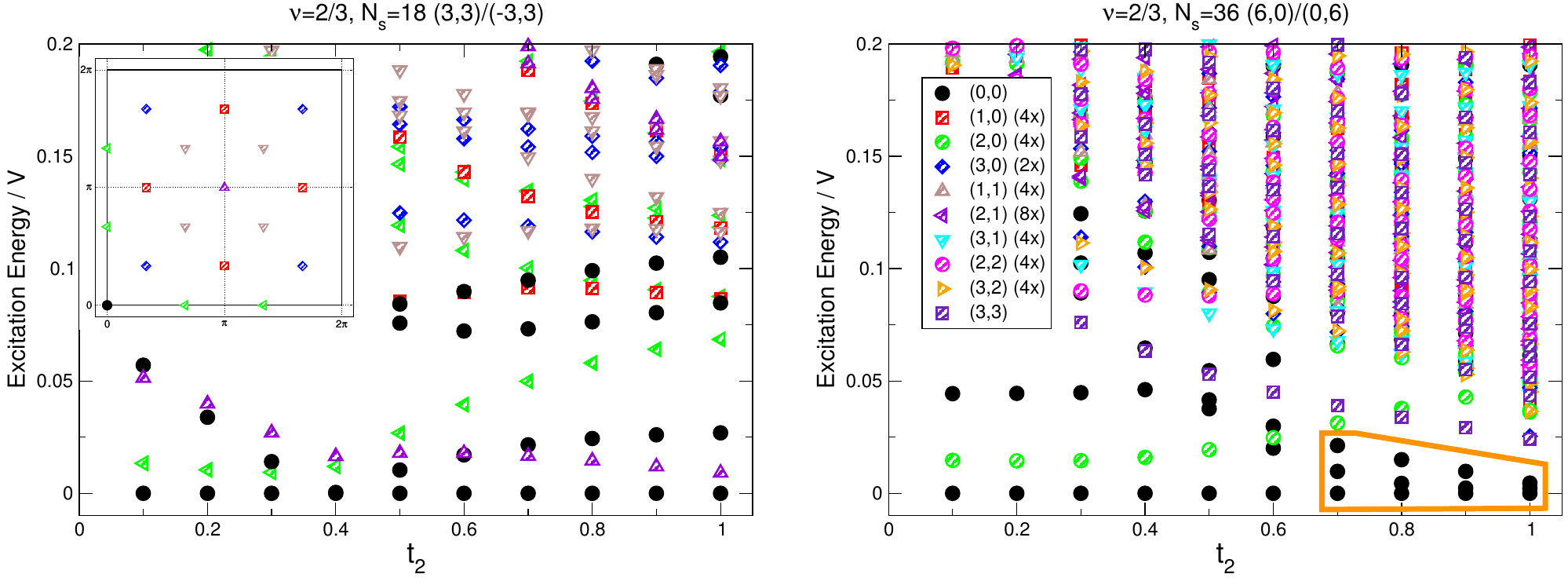}
\caption{(Color online) 
$\nu=2/3$. Unclear situation regarding a FCI state.
}
\label{fig:supp_p_2_q_3}
\end{figure*}

\begin{figure*}
\includegraphics[width=\linewidth]{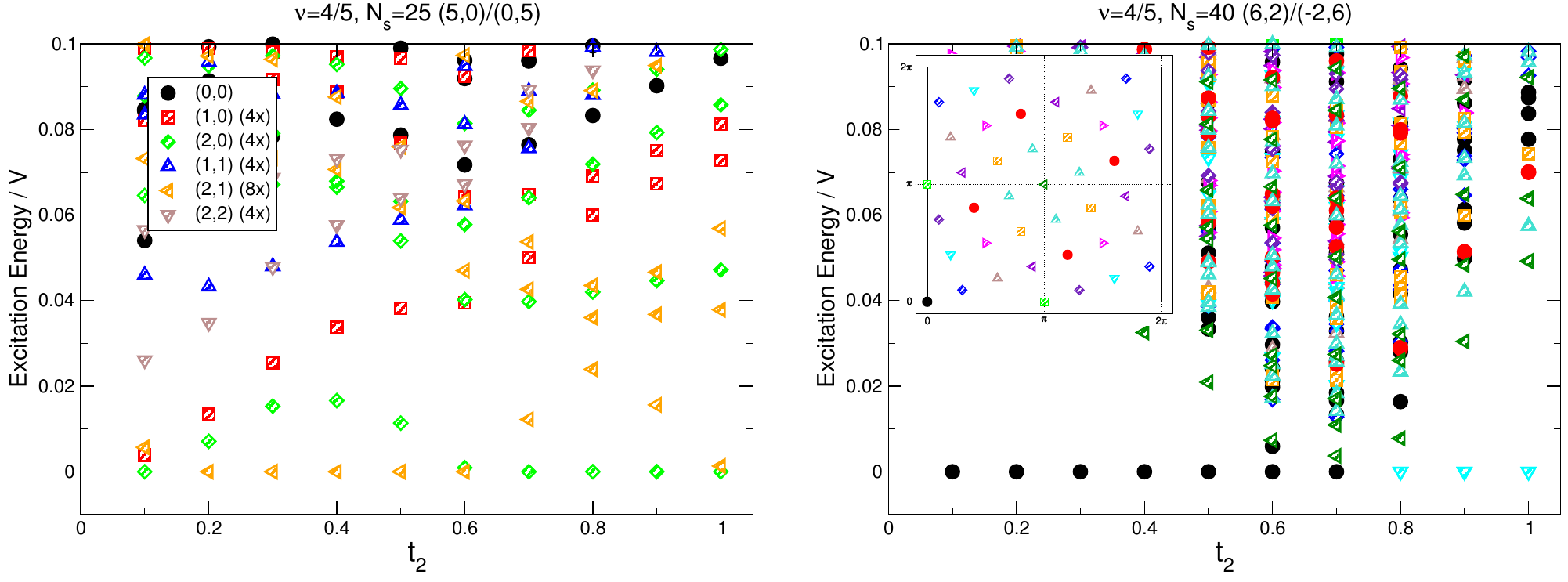}
\caption{(Color online) 
$\nu=4/5$. No evidence for a FCI state.
}
\label{fig:supp_p_4_q_5}
\end{figure*}


\begin{thebibliography}{99}
\bibitem{sp_hamilton}
$h(\mathbf k)=\sum_i d_i(\mathbf k)\sigma_i$, with $d_x(\mathbf k)=4t_1\cos(\phi)\cos(k_x/2)\cos(k_y/2)$,
$d_y(\mathbf k)=4t_1\sin(\phi) \sin(k_x/2) \sin(k_y/2)$, and $d_z(\mathbf k)=2t_2[\cos(k_x)-\cos(k_y)]$, and $\sigma_i$ are the Pauli matrices.
\end{thebibliography}
\end{document}